\documentclass[aps,prd,twocolumn,showpacs,eqsecnum,A4,amsmath,amssymb, nofootinbib]{revtex4-1}

\usepackage{graphicx}
\usepackage{multirow}

\newcommand{\dd}{{\rm{d}}} 

\def \bolde {\mbox{\boldmath$e$}}
\def \boldE {\mbox{\boldmath$E$}}
\def \boldk {\mbox{\boldmath$k$}}
\def \boldl {\mbox{\boldmath$l$}}
\def \boldm {\mbox{\boldmath$m$}}

\begin{document}

\title{A novel classification method of 2+1 spacetimes based on the Cotton scalars}

\author{J. Podolsk\'y}
\author{M. Papaj\v{c}\'ik}
\email{jiri.podolsky@mff.cuni.cz}
\email{matus.papajcik@matfyz.cuni.cz}
\affiliation{
  Charles University, Faculty of Mathematics and Physics,
  Institute of Theoretical Physics,
  V~Hole\v{s}ovi\v{c}k\'ach 2, 180~00 Prague 8, Czechia.}

\date{\today}

\begin{abstract}
A new effective approach to the algebraic classification of geometries in 2+1 gravity is presented. It uses five real Cotton scalars~$\Psi_{\rm A}$ of distinct boost weights, which are 3$D$ analogues of the Newman--Penrose scalars representing the Weyl tensor in 4$D$. The classification into types I, II, D, III, N, O is directly related to the multiplicity of the four Cotton-aligned null directions (CANDs). We derive a synoptic algorithm based on the invariants constructed from~$\Psi_{\rm A}$, and we show its agreement with the Petrov scheme based on eigenvalues and canonical Jordan form of the Cotton--York tensor. Our method is simpler and also general because it can be used in any 2+1 theory, such as Einstein's gravity or topologically massive gravity. As an example we analyze the algebraic structure of Robinson--Trautman spacetimes which include charged black holes with a cosmological constant.
\end{abstract}





\maketitle


\section{Introduction}
\label{sec:Introduction}

Einstein's general relativity (GR), conceived in 1915, is a highly successful physical theory. Formulating gravity as specific curvature of ${3+1}$ spacetime, it plays a fundamental role in contemporary astrophysics and cosmology, in particular in the theory of black holes and gravitational waves. Its predictions are in excellent agreement with various tests and observations \cite{Will:2018}.

Despite this enormous success, GR has its own limits. It predicts physical singularities inside black holes or at an initial big bang, where quantum effects become crucial. Many attempts to construct a quantum gravity theory have been made \cite{Rovelli:2010}, but it seems that none of them is yet conceptually clear, mathematically consistent, complete and (above all) experimentally tested.

In order to attack the difficult problem of quantizing gravity, various reformulations, modifications and generalizations of GR have been proposed. These include extensions to higher dimensions ${D>4}$, as in the string theory approach, but also study of lower-dimensional ${D=3}$ spacetimes. Such a \emph{2+1 gravity}, proposed in the 1960s, is nowadays a very popular research arena of fundamental physics. It provides \emph{models of quantum gravity} using various approaches \cite{Carlip:2003}, but also gives a great number of explicit \emph{exact spacetimes on a classical level} \cite{Garcia:2017}.

To classify the vast number of 2+1 spacetimes and to understand their properties, classification schemes were proposed, analogous to those in ${D=4}$ GR \cite{Stephanietal:2003}. The most important is an \emph{algebraic classification} into types I, II, D, III, N, O developed in 1954--60 by Petrov, G\'eh\'eniau, Pirani, Bell, Debever and Penrose. In its best formulation it is based on the multiplicity of 4 \emph{principal null directions} (PNDs) of the curvature Weyl tensor $C_{abcd}$ encoded in the Newman--Penrose scalars $\Psi_{\rm A}$ (${{\rm A} = 0,1,2,3,4}$) \cite{NewmanPenrose:1962, PenroseRindler:1984, Stephanietal:2003}. In 2004 this was extended to ${D>4}$ by Coley, Milson, Pravda and Pravdov\'a who introduced the concept of \emph{Weyl-aligned null directions} (WANDs), see the reviews \cite{Coley:2008, OrtaggioPravdaPravdova:2013} and \cite{KrtousPodolsky:2006} using the notation $\Psi_{\rm A}$.

The algebraic classification in ${D=3}$ was introduced by Barrow, Burd and Lancaster \cite{Barrow} and later refined by Garc\'ia, Hehl, Heinicke and Mac\'ias \cite{GHHM}. Instead of using the Weyl tensor $C_{abcd}$ (which vanishes identically), it is necessary to employ the \emph{Cotton tensor} $C_{abc}$ \cite{Cotton:1899}. In 2+1 gravity it has 5 components and can be Hodge-mapped onto the \emph{Cotton--York tensor} $Y_{ab}$ \cite{York}. It is a symmetric traceless ${3 \times 3}$ matrix, and its algebraic type can be determined by the \emph{eigenvalues} and related \emph{Jordan forms} (for details see \cite{Garcia:2017}, which also reviews another important classification method to Segre--Pleba\'nski types based on the traceless Ricci tenor).

Here we propose a \emph{novel, practical method} of algebraic classification in 2+1 gravity, consistent with \cite{GHHM}. Assuming no field equations, it is also more \emph{general}. We  directly use the Cotton scalars~$\Psi_{\rm A}$ which are the null-triad projections of $C_{abc}$. These are the ${D=3}$ counterparts of the Newman--Penrose Weyl scalars of GR, and determine the multiplicity of the CANDs --- in analogy with PNDs (in ${D=4}$) and WANDs (in ${D>4}$).


\section{Cotton scalars~$\Psi_{\rm A}$}
\label{sec:Cotton}

In a general 3$D$ spacetime with the metric $g_{ab}$ of signature $(-,+,+)$, the curvature is given by the \emph{Ricci tensor} $R_{ab}$, \emph{Ricci scalar} ${R \equiv {R_a}^{a}}$ and \emph{Cotton tensor} \cite{Garcia:2017}, Ch.~20,
\begin{equation} \label{Cotton_Definition}
C_{abc} \equiv 2\, \Big( \nabla_{[a}R_{b]c}-\tfrac{1}{4}\nabla_{[a}R\,g_{b]c} \Big),
\end{equation}
where $\nabla$ is the metric connection. It is antisymmetric (${C_{(ab)c} = 0 = C_{[abc]}}$) and traceless
(${{C_{ab}}^{a} = 0}$), so that it has \emph{5 independent components}.
The Cotton tensor $C_{abc}$ plays the role similar to the Weyl tensor $C_{abcd}$ representing a free gravitational field in Einstein's 4$D$ gravity. In the Newman--Penrose formalism \cite{NewmanPenrose:1962, PenroseRindler:1984, Stephanietal:2003} the Weyl tensor is encoded in 5 complex scalars~$\Psi_{\rm A}$, defined as projections of $C_{abcd}$ onto specific combinations of some \emph{null tetrad} vectors. Since $C{_{abcd} \equiv 0}$ in any 2+1 geometry, it cannot be used for classification in 3$D$. However, one can introduce the \emph{null triad} basis ${\{ \bolde_a \}\equiv\{ \boldk, \, \boldl, \, \boldm \}}$ normalized~as
\begin{equation} \label{null-comp}
k_a \, l^a=-1 \, , \qquad m_a \, m^a=1 \, ,
\end{equation}
${k_a \, k^a = 0 = l_a \, l^a}$ and ${k_a \, m^a = 0 = l_a \, m^a}$. It means that $\boldk$ and $\boldl$ are \emph{null vectors}, while $\boldm$ is the \emph{spatial unit vector} orthogonal to both $\boldk$ and $\boldl$.
Next we define the Newman--Penrose-type real \emph{Cotton scalars}~$\Psi_{\rm A}$ as the projections
\begin{align}
\Psi _0 &\equiv C_{abc} \, k^a \, m^b \, k^c \, , \nonumber\\ 
\Psi _1 &\equiv C_{abc} \, k^a \, l^b \, k^c \, , \nonumber\\
\Psi _2 &\equiv C_{abc} \, k^a \, m^b \, l^c \, , \label{Psi}\\
\Psi _3 &\equiv C_{abc} \, l^a \, k^b \, l^c \, , \nonumber\\
\Psi _4 &\equiv C_{abc} \, l^a \, m^b \, l^c \, . \nonumber   
\end{align}
Equivalently,
$\Psi_1 =C_{abc} \, k^a \, m^b \, m^c$, $\Psi_3 =C_{abc} \, l^a \, m^b \, m^c$,
$\Psi_2 =C_{abc} \, m^a \, l^b \, k^c  =\tfrac{1}{2}C_{abc} \, k^a \, l^b \, m^c$.


\section{The classification based on $\Psi_{\rm A}$}
\label{sec:PsiAclassification}

The specific \emph{algebraic types} of 2+1 geometries can now be defined by very simple conditions, namely that \emph{in a suitable null triad} ${\{ \boldk, \, \boldl, \, \boldm \}}$ some of the \emph{Cotton scalars~$\Psi_{\rm A}$ vanish}, as given in Tab.~\ref{Tab-classification}.

\vspace{-3.0mm}
\begin{table}[!h]
\begin{center}
\caption{\label{Tab-classification} A simple algebraic classification of 2+1 geometries.}
\vspace{2.0mm}
\begin{tabular}{c|clll}
\hline
\hline
type && \qquad\qquad the conditions\\[2pt]
\hline
I   && ${\Psi _0=0}$\,,                              && $\Psi _1\ne0$ \\
II  && ${\Psi _0=\Psi _1=0}$\,,                      && $\Psi _2\ne0$ \\
III && ${\Psi _0=\Psi _1=\Psi _2=0}$\,,              && $\Psi _3\ne0$ \\
N   && ${\Psi _0=\Psi _1=\Psi _2=\Psi _3=0}$\,,      && $\Psi _4\ne0$ \\
D   && ${\Psi _0=\Psi _1 = 0 = \Psi _3=\Psi _4}$\,,  && $\Psi _2\ne0$ \\
O   && all ${\Psi_{\rm A}=0}$  && \\[1pt]
\hline
\hline
\end{tabular}
\end{center}
\end{table}

\noindent
In fact, formally these are the same conditions as for the ${D=4}$  classification related to the multiplicity of the PNDs of the Weyl tensor, see Sec.~4.3 of \cite{Stephanietal:2003}, or the multiplicity of the WANDs in ${D>4}$ \cite{OrtaggioPravdaPravdova:2013}.


\section{Classification invariants}
\label{invariants}

A \emph{general choice} of the null triad ${\{ \boldk, \, \boldl, \, \boldm \}}$ does not provide vanishing Cotton scalars as prescribed in Tab.~\ref{Tab-classification}. However, for each algebraic type the required canonical forms of~$\Psi_{\rm A}$ can be achieved by a suitable Lorentz transformation. This also leads to the important concept of Cotton-aligned null direction (CAND), see Sec.~\ref{CAND}.

\begin{figure}[!t]
\begin{center}
\includegraphics[scale=0.6]{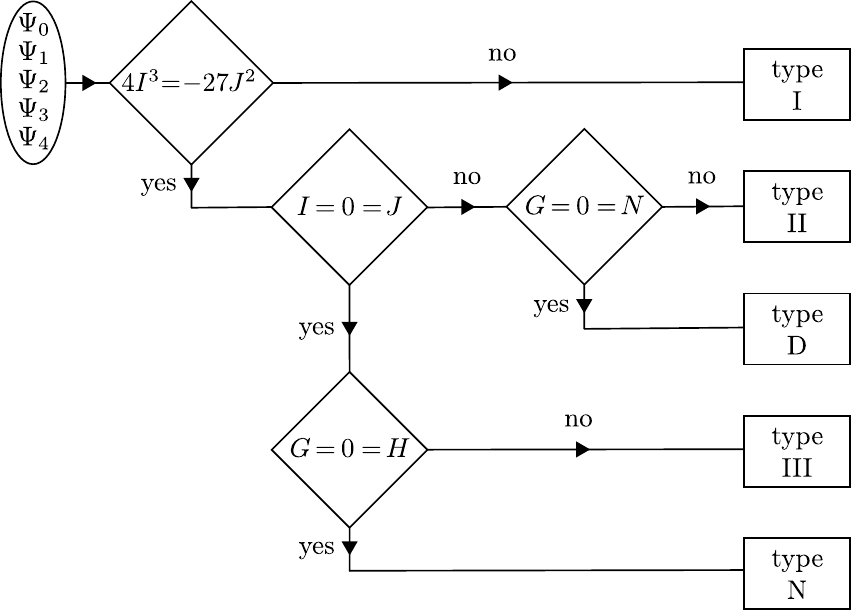}
\vspace{-4.0mm}
\caption{
Flow diagram for determining the algebraic type of a 2+1 geometry using the invariants~\eqref{Invariants_IJGHN} constructed from the Cotton scalars~$\Psi_{\rm A}$ (if ${\Psi_4 \neq 0}$). Type~O is a conformally flat spacetime with vanishing Cotton tensor. Application of the diagram on various explicit geometries can be found in \cite{PapajcikPodolsky:2023}.}
\label{Flow_diagram}
\end{center}
\end{figure}

In practise it is not necessary to find the null triad with the canonical forms of~$\Psi_{\rm A}$. It can be proven \cite{PapajcikPodolsky:2023} that the algebraic type of a given 2+1 geometry can be determined using the Cotton scalars \eqref{Psi} evaluated \emph{in any null triad} satisfying \eqref{null-comp}. Such a classification can be done using the \emph{scalar polynomial invariants}
\begin{align}
I & \equiv \Psi _0\Psi _4 -2\,\Psi _1\Psi _3 -3\,\Psi _2^2 \, ,\nonumber\\
J & \equiv 2\,\Psi _0\Psi _2\Psi _4 +2\,\Psi _1\Psi _2\Psi _3 +2\,\Psi _2^3 + \Psi _0\Psi _3^2 - \Psi _4 \Psi _1^2  \, , \nonumber \\
G &\equiv \Psi _1\Psi _4^2 -3\,\Psi _2\Psi _3\Psi _4 -\Psi _3^3 \, ,\label{Invariants_IJGHN}\\
H &\equiv 2\,\Psi _2\Psi _4 + \Psi _3^2 \, , \nonumber\\
N &\equiv 3\, H^2 + \Psi _4^2\, I \, .\nonumber
\end{align}
In fact, ${I = \tfrac{1}{4}\, C_{abc}\, C^{abc}}$ and
${J = \tfrac{1}{6}\, C_{abc}\, C^{abd}\, {Y^c}_d}$, where $Y_{ab}$ is the Cotton--York tensor \eqref{Cotton-York}.
Useful \emph{algorithm of algebraic classification} is given by the flow diagram in Fig.~\ref{Flow_diagram} (it is an analogue of the ${D=4}$ diagram presented in Fig.~9.1 of \cite{Stephanietal:2003}). The procedure is \emph{not applicable if} ${\Psi_4 = 0}$. In this case, when ${\Psi_0 \neq 0}$ we can perform a swap ${\Psi_0 \leftrightarrow -\Psi_4}$, ${\Psi_1 \leftrightarrow - \Psi_3}$, after which Fig.~\ref{Flow_diagram} can be used. When ${\Psi_4 =0 =\Psi_0}$, it is necessary to employ~Tab.~\ref{Tab:Algorithm-special-case}.

\vspace{-3.0mm}
\begin{table}[!h]
\begin{center}
\caption{\label{Tab:Algorithm-special-case} Algebraic classification of 2+1 geometries for the special case ${\Psi_4 =0 =\Psi _0}$.}
\vspace{2.0mm}
\begin{tabular}{c|c|c|c|l}
\hline
\hline\\[-12pt]
& \multicolumn{1}{c|}{} & \multicolumn{2}{c|}{}\\[-13pt]
\multirow{5}{*}{${\Psi _1 =0}$}  & \multirow{2}{*}{${\Psi _2 =0}$} & \multicolumn{2}{c|}{${\Psi _3 =0}$}
& type O\\[-6pt]
& \multicolumn{1}{c|}{} & \multicolumn{2}{c|}{}\\[-8pt]
                              &                               & \multicolumn{2}{c|}{$\Psi _3 \neq 0$} & type III\\
                               \cline{2-5}
& \multicolumn{1}{c|}{} & \multicolumn{2}{c|}{}\\[-12pt]
                              & \multirow{2}{*}{$\Psi _2 \neq 0$} & \multicolumn{2}{c|}{$\Psi _3 =0$} & type D\\[-6pt]
& \multicolumn{1}{c|}{} & \multicolumn{2}{c|}{}\\[-8pt]
                              &                               & \multicolumn{2}{c|}{$\Psi _3 \neq 0$} & type II \\[0pt] \hline
& \multicolumn{1}{c|}{} & \multicolumn{2}{c|}{}\\[-12pt]
\multirow{7}{*}{$\Psi _1 \neq 0$} & \multirow{2}{*}{$\Psi _2 =0$} & \multicolumn{2}{c|}{$\Psi _3 =0$} & type III\\[-6pt]
& \multicolumn{1}{c|}{} & \multicolumn{2}{c|}{}\\[-8pt]
                                  &                               & \multicolumn{2}{c|}{$\Psi _3 \neq 0$} & type I\\[0pt] \cline{2-5}
& \multicolumn{1}{c|}{} & \multicolumn{2}{c|}{}\\[-12pt]
                                  & \multirow{4}{*}{$\Psi _2 \neq 0$} & \multicolumn{2}{c|}{$\Psi _3 =0$} & type II\\[0pt] \cline{3-5}
& \multicolumn{1}{c|}{} & \multicolumn{1}{c|}{} & \multicolumn{1}{c|}{}\\[-11pt]
                                  &                               & \multirow{2}{*}{$\Psi _3 \neq 0$} & $9\Psi                         																			 _2^2 =-8\Psi _1\Psi _3$ & type II\\[-2pt]
& \multicolumn{1}{c|}{} & \multicolumn{1}{c|}{} & \multicolumn{1}{c|}{}\\[-11pt]
                                  &                               &                                   & $9\Psi                         																			 _2^2 \neq -8\Psi _1\Psi _3$ & type I \\[1pt]
\hline
\hline
\end{tabular}
\end{center}
\end{table}


\section{Cotton-aligned null direction (CAND)}
\label{CAND}

The Cotton scalars~$\Psi_{\rm A}$ depend on the choice of the null triad ${\{ \boldk, \, \boldl, \, \boldm \}}$. However, as in ${D \ge 4}$ this freedom is simply given by \emph{Lorentz transformations} at a given point of the spacetime. There are 3 subgroups preserving~\eqref{null-comp}:
\begin{align}
\boldk' & = B\,\boldk \, ,\quad
  \boldl' = B^{-1}\, \boldl  \, ,\quad\,
  \boldm' = \boldm \, , \label{boost} \\
\boldk' & = \boldk \, , \quad
  \boldl' = \boldl + \sqrt2\, L\,\boldm + L^2\, \boldk  \, ,\quad\,
  \boldm' = \boldm + \sqrt2\, L\,\boldk \, , \label{kfixed} \\
\boldk' & = \boldk + \sqrt2\, K\,\boldm + K^2\, \boldl  \, ,\quad
  \boldl' = \boldl \, ,\quad
  \boldm' = \boldm + \sqrt2\, K\,\boldl  \, ,\label{lfixed}
\end{align}
where ${B, K, L}$ are \emph{real parameters}. Under the \emph{boost} \eqref{boost}, ${\Psi'_{\rm A} = B^{2-{\rm A}}\,\Psi_{\rm A}}$, i.e. $\Psi_{\rm A}$  in \eqref{Psi} are \emph{ordered according to their specific boost weights}, which is the corresponding \emph{power}~${(2-{\rm A})}$ of the boost parameter~$B$. Under the \emph{null rotation \eqref{kfixed} with fixed} ${\boldk=\boldk'}$, the scalars transform as ${\Psi_0' = \Psi_0}$, ${\Psi_1' = \Psi_1 + \sqrt{2}\,L\,\Psi_0}$, etc. \cite{PapajcikPodolsky:2023}. The classification in Tab.~\ref{Tab-classification} is thus \emph{invariant with respect to both \eqref{boost} and \eqref{kfixed}}. The \emph{null rotation \eqref{lfixed} with fixed} ${\boldl=\boldl'}$ transforms the Cotton scalars \eqref{Psi} as
\begin{align}\label{eq:null-rotatin-fixed-l}
\Psi_0' &= \Psi_0+2\sqrt{2}\,K\,\Psi_1 +6K^2\,\Psi_2 -2\sqrt{2}\,K^3\,\Psi_3 - K^4\,\Psi_4 \, ,
\end{align}
etc. Now, a crucial observation is that \eqref{eq:null-rotatin-fixed-l} \emph{always} allows us to \emph{achieve}~${\Psi'_0=0}$ by a \emph{suitable choice} of the parameter $K$, so that in the new triad  ${\{ \boldk', \, \boldl', \, \boldm' \}}$ the condition for type~I given in Tab.~\ref{Tab-classification} is satisfied. Such a frame can be called the \emph{principal null triad}, and we name its null vector $\boldk'$ the \emph{Cotton-aligned null direction}, abbreviated as CAND. A CAND is the 2+1 analogue of a PND (principal null direction) of the Weyl tensor in ${D=4}$ GR, and of a WAND (Weyl-aligned null direction) in  ${D\ge4}$  gravity \cite{Stephanietal:2003, OrtaggioPravdaPravdova:2013}.

This also proves that \emph{all 2+1 geometries with ${C_{abc}\ne0}$ are of algebraic type~I, or more special}. This is also true for ${3+1}$ geometries, considering the Weyl tensor instead of the Cotton tensor.

Moreover, the CAND can be explicitly found. In view of \eqref{eq:null-rotatin-fixed-l}, the condition~${\Psi'_0=0}$ reads
\begin{equation} \label{Class_eq}
\Psi_4\,K^4 + 2\sqrt{2}\,\Psi_3\,K^3 - 6\,\Psi_2\,K^2 - 2\sqrt{2}\,\Psi_1\,K  - \Psi_0 =0 \, .
\end{equation}
It is an algebraic equation of the 4th order for $K$, which admits \emph{four complex solutions}  (not necessarily distinct). Therefore, at any event of the 2+1 spacetime \emph{there exist four CANDs} determined by the local algebraic structure of the Cotton tensor. Each of these four CANDs~$\boldk'$ is obtained by \eqref{lfixed} where the parameter $K$ is the root of \eqref{Class_eq}, although it is generally difficult to find them explicitly. Multiplicity of the roots $K$ thus implies the same \emph{multiplicity of the CANDs}. These are uniquely related to the algebraic types, as we discuss next.


\section{CANDs multiplicity, complexity, and the algebraic (sub)types}
\label{CAND-multiplicity}

A 2+1 spacetime is said to be \emph{algebraically general} if its CANDs, i.e. the four roots of \eqref{Class_eq}, are \emph{all distinct}. Such a spacetime is of algebraic \emph{type~I}. A spacetime is \emph{algebraically special} if at least two of its CANDs \emph{coincide}. If \emph{just two} CANDs~$\boldk$ coincide, it is of \emph{type~II}. Higher multiplicity defines \emph{type~III} (triple CAND/root) and the most special \emph{type~N} (quadruple CAND/root) geometries. In addition, there is a degenerate case \emph{type~D} --- a subtype of type~II with two distinct CANDs~$\boldk$ and~$\boldl$, both of multiplicity 2 (\emph{two pairs} of coinciding roots), see Tab.~\ref{Tab:algebraic-types}.

\begin{table}[!h]
\begin{center}
\caption{\label{Tab:algebraic-types} Algebraic types of 2+1 geometries are uniquely related to the multiplicity of the Cotton-aligned null directions (CANDs), i.e. to the multiplicity of the four roots of the key equation \eqref{Class_eq}. The last column contains the roots $K$ and $L$ for the canonical Cotton scalars~$\Psi_{\rm A}$ given in Tab.~\ref{Tab-classification}.}
\vspace{2.0mm}
\begin{tabular}{cccll}
\hline
\hline
\\[-10pt]
  type & CANDs & {multiplicity} & & \hspace{-5mm} canonical roots \\[2pt]
\hline
\\[-8pt]
   I
   & \hbox{
   \rotatebox[origin=c]{-30}{$\leftarrow$}\hspace{-3mm}
   \raisebox{1.5mm}{\rotatebox[origin=c]{-60}{$\leftarrow$}}\hspace{-1mm}
   \raisebox{1.5mm}{\rotatebox[origin=c]{60}{$\rightarrow$}}\hspace{-3mm}
   \rotatebox[origin=c]{30}{$\rightarrow$}}
   & \hspace{-0mm}${1+1+1+1}$\
   & \hspace{1.4mm}${K=0}$,
   & \hbox{3 other distinct roots} \\[2pt]
   II
   & \hbox{
   \rotatebox[origin=c]{-30}{$\leftarrow$}\hspace{-3mm}
   \raisebox{1.5mm}{\rotatebox[origin=c]{-60}{$\leftarrow$}}\hspace{-1mm}
   \raisebox{0.6mm}{\rotatebox[origin=c]{45}{$\Rightarrow$}}}
   & ${1+1+2}$
   & ${K^2=0}$,
   & \hbox{2 other distinct roots} \\[6pt]
   D
   & \hbox{
   \rotatebox[origin=c]{-45}{$\Leftarrow$}\,\rotatebox[origin=c]{45}{$\Rightarrow$}}
   & ${2+2}$
   & ${K^2=0}$,
   & ${L^2=0}$    \\[4pt]
   III
   & \hbox{
   \rotatebox[origin=c]{-30}{$\leftarrow$}\hspace{-1mm}
   \raisebox{0.6mm}{\rotatebox[origin=c]{45}{$\Rrightarrow$}}}
   & ${1+3}$
   & ${K^3=0}$,
   & \hspace{1.5mm}${L=0}$\  or\  ${K \ne 0}$ \\[6pt]
   N
   & {\Large
   \hbox{
   \rotatebox[origin=c]{45}{$\Rightarrow$}\hspace{-6.1mm}
   \raisebox{-0.30mm}{\rotatebox[origin=c]{45}{$\Rightarrow$}}}
   }
   & $4$
   & ${K^4=0}$
   & \\[1pt]
\hline
\hline
\end{tabular}
\end{center}
\end{table}

If $\boldk$ is a CAND then ${\Psi_0=0}$, corresponding to the root ${K=0}$ of  \eqref{Class_eq}. Type~II arises when ${\Psi_1=0}$, type~III  when ${\Psi_1=\Psi_2=0}$, and type~N when ${\Psi_1=\Psi_2=\Psi_3=0}$, in which case ${K^4=0}$ gives the quadruple CAND~$\boldk$. For type~D spacetimes with ${\Psi_0=\Psi_1=0=\Psi_3=\Psi_4}$, Eq.~\eqref{Class_eq} reduces to ${\Psi_2\, K^2 = 0}$, so that $\boldk$ is the double CAND. Moreover (swapping ${\boldk \leftrightarrow \boldl}$, so that ${\Psi_0 \leftrightarrow \Psi_4 }$, ${\Psi_1 \leftrightarrow \Psi_3}$, ${\Psi_2 \leftrightarrow -\Psi_2 }$), the vector $\boldl$ is another double CAND, and the condition ${\Psi_4=\Psi_3=0}$ implies ${L^2  =0}$.

Such a classification in 2+1 gravity is actually more subtle because the key \emph{real} equation \eqref{Class_eq} can have some \emph{complex roots} $K$. Some of the vectors~$\boldk$ representing CANDs can thus \emph{formally be complex}. This cannot happen in 3+1 gravity or in ${D>4}$. It is thus natural to suggest a \emph{subclassification}:

\begin{itemize}
\item subtypes~I$_{\rm r}$, II$_{\rm r}$ and D$_{\rm r}$: \emph{all four (possibly multiple) CANDs are real},

\item subtypes~I$_{\rm c}$, II$_{\rm c}$ and D$_{\rm c}$: \emph{some of the CANDs are complex}.
\end{itemize}

\noindent
It can be shown that type {III\,$\equiv$\,III$_{\rm r}$} and type {N\,$\equiv$\,N$_{\rm r}$}.

Explicit conditions for these subtypes, the relation to Class~$\text{I}^{\prime }$ geometries with complex eigenvalues of the Cotton--York tensor, and the Petrov--Segre types I$_{\mathbb R}$ and I$_{\mathbb C}$ in topologically massive gravity will be presented in \cite{PapajcikPodolsky:2023}, together with the Bel--Debever criteria for all types.

\begin{table*}[!t]
\begin{center}
\caption{\label{Tab:Normal-Types} Algebraic classification based on the Jordan normal forms
and eigenvalues $\lambda_i$ of the Cotton--York tensor~${Y_a}^b$, the corresponding special values of the Cotton scalars~$\Psi_{\rm A}$, and the scalar invariants. It shows the equivalence of these classifications.}
\vspace{2.0mm}
\begin{tabular}{cccc}
\hline
\hline
\\[-8pt]
   algebraic type & \qquad Jordan normal form of ${Y_a}^b$ \qquad &  special Cotton scalars & invariants \\[2pt]
\hline
\\[-4pt]
   I
   & $\begin{pmatrix} \lambda _1 & 0 & 0\\ 0 & \lambda _2 & 0\\ 0 & 0 & -\lambda _1 -\lambda _2 \end{pmatrix}$
   & $\begin{matrix} \Psi _1 =0=\Psi _3  \\ \quad \Psi _0 =\frac{1}{2}(\lambda _1-\lambda _2)=-\Psi _4 \quad \\ \Psi _2=\frac{1}{2}(\lambda _1+\lambda _2) \end{matrix}$
   & $\begin{matrix} \quad I=\lambda _1\lambda _2 - (\lambda_1 + \lambda_2)^2 \\ \quad J=\lambda _1\lambda _2\,(\lambda _1+\lambda _2) \end{matrix}$
   \\[20pt]
   II
   & $\begin{pmatrix} \lambda _1 -1 & -1 & 0\\ 1 & \lambda _1 +1 & 0\\ 0 & 0 & -2\lambda _1 \end{pmatrix}$
   & $\begin{matrix} \Psi _0 = 0\,, \ \Psi _1 =0 = \Psi _3\\ \Psi _2=\lambda _1 \\ \Psi _4 =2 \end{matrix}$
   & $\begin{matrix} I=-3\,\lambda_1^2  \\ J=2\,\lambda_1^3 \\  G=0\,,\ N = 36\,\lambda_1^2 \end{matrix}$
   \\[26pt]
   D
   & $\begin{pmatrix} \lambda _1 & 0 & 0\\ 0 & \lambda _1 & 0\\ 0 & 0 & -2\lambda _1 \end{pmatrix}$
   & $\begin{matrix} \Psi _0 =0=\Psi _4\\ \Psi _1 =0=\Psi _3 \\ \Psi _2=\lambda _1\end{matrix}$
   & $\begin{matrix} I=-3\,\lambda_1^2  \\ J=2\,\lambda_1^3 \\  G=0=N \end{matrix}$
   \\[20pt]
   III
   & $\begin{pmatrix} 0 & 0 & 1\\ 0 & 0 & -1\\ -1 & -1 & 0 \end{pmatrix}$
   & $\begin{matrix} \Psi _0 =0=\Psi _4 \\ \Psi _1 =0=\Psi _2 \\ \Psi _3=\sqrt{2} \end{matrix}$
   & $\begin{matrix} I=0=J \\ G=-2\sqrt{2}\,,\  H=2 \end{matrix}$
   \\[20pt]
   N
   & $\begin{pmatrix} -1 & -1 & 0\\ 1 & 1 & 0\\ 0 & 0 & 0 \end{pmatrix}$
   & $\begin{matrix} \Psi _0 =0= \Psi _2 \\ \Psi _1 =0=\Psi _3\\ \Psi _4=2\end{matrix}$
   & $\begin{matrix} I=0=J \\ G=0=H\\ \end{matrix}$
   \\[20pt]
\hline
\hline
\end{tabular}
\end{center}
\end{table*}



\section{Equivalence with classification based on the Cotton--York tensor}
\label{sec:CottonYork}

In previous studies, the Hodge dual of the Cotton tensor $C_{abc}$ was employed. The \emph{Cotton--York tensor} \cite{York} is defined as ${Y_{ab}\equiv \bolde_a \rfloor\, ^*\mathbf{C}_b }$, where  $\mathbf{C}_b$ is the Cotton 2-form ${\mathbf{C}_b \equiv \tfrac{1}{2}\,C_{m n b} \, \boldsymbol{\omega}^m \wedge \boldsymbol{\omega}^n}$, see Eq.~(20.111) in \cite{Garcia:2017}. Explicitly,
\begin{align} \label{Cotton-York}
Y_{ab} &=\tfrac{1}{2} \, g_{ak}\, \omega^{kmn}  \, C_{mnb} \, ,
\end{align}
where the Levi-Civita tensor reads ${\omega _{abc}=-\sqrt{-g} \,\, \varepsilon _{abc}}$. It has 5 components because it is symmetric and traceless (${Y_{ab}=Y_{ba}}$ and ${{Y_{a}}^{a}=0}$).
It can be shown \cite{PapajcikPodolsky:2023} that the \emph{general Cotton--York tensor} in the triad \eqref{null-comp} is
\begin{align} \label{York}
Y_{ab}=&-\Psi _0 \, l_a \, l_b +\Psi _1 (l_a \, m_b +m_a \, l_b)  \nonumber \\
& -\Psi _2 (l_a \, k_b + k_a \, l_b + 2 \, m_a \, m_b )\nonumber \\
& -\Psi _3 (k_a \, m_b + m_a \, k_b)+\Psi _4 \, k_a \, k_b \, .
\end{align}
This can be used to show the equivalence of our new method of classification, based on the Cotton scalars $\Psi_{\rm A}$ and the multiplicity of CANDs, with the ``Petrov'' scheme based on the Jordan form and eigenvalues of the Cotton--York tensor \cite{Barrow, GHHM, Garcia:2017}. To this end, we express $Y_{ab}$ in an \emph{orthonormal basis} ${\{ \boldE_0, \, \boldE_1, \, \boldE_2 \}}$ associated with ${\{ \boldk, \, \boldl, \, \boldm \}}$ via
${\boldE_0\equiv\tfrac{1}{\sqrt{2}}(\boldk+\boldl)}$,
${\boldE_1\equiv\tfrac{1}{\sqrt{2}}(\boldk-\boldl)}$,
${\boldE_2\equiv\boldm}$, so that in this basis the metric reads ${g_{ab}= \text{diag}(-1,1,1)}$.
Using \eqref{York} we calculate all orthonormal projections (such as ${Y_{00}\equiv E_{0}^{\,a} E_{0}^{\,b}\,Y_{ab}}$ etc.), and evaluate ${{Y_a}^b \equiv Y_{ac} \, g^{cb}}$ as
\begin{align}\label{class}
{Y_a}^b=
\begin{pmatrix}
{\displaystyle \Psi _2 + \frac{\Psi _0 - \Psi _4}{2}  }
&
{\displaystyle -\frac{\Psi _0 + \Psi _4}{2}}
&
{\displaystyle -\frac{\Psi _1 - \Psi _3}{\sqrt{2}}}
\\[4mm]
{\displaystyle \frac{\Psi _0 + \Psi _4}{2}}
&
{\displaystyle \Psi _2 - \frac{\Psi _0 - \Psi _4}{2} }
&
{\displaystyle -\frac{\Psi _1 + \Psi _3}{\sqrt{2}}}
\\[4mm]
{\displaystyle \frac{\Psi _1 - \Psi _3}{\sqrt{2}}}
&
{\displaystyle -\frac{\Psi _1 + \Psi _3}{\sqrt{2}}}
&
-2\Psi _2
\end{pmatrix} .
\end{align}
This matrix is traceless but \emph{not symmetric}, so the roots of the characteristic polynomial ${\det \,({Y_a}^b - \lambda\,{\delta_a}^b ) = 0}$ may be \emph{complex}. According to the \emph{eigenvalues} $\lambda _1, \lambda _2$  and ${\lambda_3 = - \lambda _1 - \lambda _2}$, the \emph{Petrov types} are defined such that the case ${\lambda _1 \neq \lambda _2}$ gives type~I, ${\lambda _1 = \lambda _2 \neq 0}$ gives type~II or type~D, while ${\lambda _1 = \lambda _2 = 0 = \lambda _3}$ gives type~III or type~N.

We can find the \emph{Jordan normal forms} defining the algebraic types of 2+1 geometries, as presented in Tab.~\ref{Tab:Normal-Types}. For types~I and~D the first two columns are a copy of Tab.~1.2.1 of \cite{Garcia:2017}. For types~II and~N we performed a \emph{similarity transformation} between the Jordan form~$J$ presented in \cite{Garcia:2017} and the \emph{normal form}~${A \, J \, A^{-1}}$ with
${A=}${\tiny ${\begin{pmatrix} -1 & 1 & 0\\ 1 & 0 & 0\\ 0 & 0 & 1 \end{pmatrix}}$}. For type~III we employed
${A=}${\tiny ${\begin{pmatrix} -1 & 0 & 1\\ 1 & 0 & 0\\ 0 & -1 & 0\end{pmatrix}}$}.
Such forms of~${Y_a}^b$ can be uniquely identified with the special values of the Cotton scalars~$\Psi_{\rm A}$  using \eqref{class}, see the third column of Tab.~\ref{Tab:Normal-Types}.
It confirms that the ``Petrov types'' defined by the Jordan forms of~${Y_a}^b$ are equivalent to our approach. Indeed, evaluating the invariants~\eqref{Invariants_IJGHN} for the special values of~$\Psi_{\rm A}$ (the last column in Tab.~\ref{Tab:Normal-Types}) the algorithm in Fig.~\ref{Flow_diagram}  \emph{gives the same algebraic types} (for D and III  we must employ Tab.~\ref{Tab:Algorithm-special-case} because ${\Psi_0 =0 =\Psi_4}$). Our method also shows the unique relation to multiplicity of CANDs, in full analogy with PNDs in ${D=4}$ gravity, see Sec.~4.3 of \cite{Stephanietal:2003}, and WANDs in ${D>4}$ theories \cite{OrtaggioPravdaPravdova:2013}.


\section{Robinson--Trautman spacetimes}
\label{sec:example}

The usefulness of the new classification method can be demonstrated on treating a class of 2+1 Robinson--Trautman spacetimes with a \emph{cosmological constant}~$\Lambda$ and an \emph{aligned electromagnetic field}, which include black holes. In \cite{PodolskyPapajcik:2022} we derived their general form
\begin{align}\label{RTmetric-final}
\dd s^2 &= \frac{r^2}{P^2} \big( \dd x + e \,P^2 \dd u \big)^2 -2\,\dd u\,\dd r - 2H \, \dd u^2\, ,\\
&  2H = -m + \kappa_0\,Q^2 \ln \Big|\frac{Q}{r}\Big| - 2\,(\ln Q)_{,u}\,r - \Lambda\,r^2 \,, \nonumber
\end{align}
with the Maxwell field potential ${{\mathbf{A}} =  Q \,\ln (r/r_0)\,  \dd u}$, see Eqs.~(180), (182) of \cite{PodolskyPapajcik:2022}. Here ${m \equiv \mu\, Q^2}$, $\mu$ is a constant and $Q(u)$ is \emph{any} function of $u$, while the functions $P(u,x)$, $e(u,x)$ satisfy the Einstein field equation ${(Q/P)_{,u} = Q\,(e\,P)_{,x}}$. The Cotton tensor \eqref{Cotton_Definition} of the solution \eqref{RTmetric-final} has components
${C_{urr} = \tfrac{1}{2}\kappa_0\, Q^2\,r^{-3}}$,
${C_{xrx} = \tfrac{1}{2}\kappa_0\,(Q^2/P^2)\, r^{-1}}$,
${C_{xru} = C_{urx} = \tfrac{1}{2}\kappa_0 \,e\, Q^2\, r^{-1}}$,
and much more complicated expressions for  $C_{ruu}$, $C_{uxu}$ and $C_{uxx}$. However, using the null triad
${\boldk = \partial_r}$, ${\boldl = \partial_u - H\,\partial_r - e\,P^2\,\partial_x}$,
${\boldm = (P/r)\,\partial_x}$, definition \eqref{Psi} yields \emph{very simple Cotton scalars}
\begin{align}\label{eq:null-rotatin-fixed-k-RT-EM}
\Psi_0 &= 0 = \Psi_4\, , \quad
 \Psi_2 = 0\, , \\[1mm]
\Psi_1 &=  -\frac{\kappa_0 Q^2}{2r^3}  \, , \quad
 \Psi_3 = \frac{1}{2}\Big( \,m -\kappa_0 Q^2 \ln \Big| \frac{Q}{r} \Big| +\Lambda\,r^2 \, \Big)\,\Psi_1
\, . \nonumber
\end{align}

Because ${\Psi_0=0}$ and generally ${\Psi_1\ne0}$, Tab.~\ref{Tab-classification} indicates that such spacetimes are \emph{of algebraic type~I}. Moreover, it follows from Secs.~\ref{CAND},~\ref{CAND-multiplicity} that  ${\boldk=\partial_r}$ is the CAND. It coincides with the null direction of the aligned electromagnetic field. Because ${\Psi _4=0}$, the second distinct CAND is ${\boldl}$.

All scalars \eqref{eq:null-rotatin-fixed-k-RT-EM} vanish when ${Q=0}$, corresponding to \emph{vacuum} solutions with $\Lambda$, and thus (anti-)de~Sitter or Minkowski spaces which are \emph{conformally flat} (type~O). In the case ${Q\ne0}$ with an aligned electromagnetic field the main invariants \eqref{Invariants_IJGHN} are ${I = -2\,\Psi_1\Psi_3}$ and ${J=0}$. Because ${\Psi_4 =0 =\Psi_0}$, we must employ Tab.~\ref{Tab:Algorithm-special-case} instead of Fig.~\ref{Flow_diagram}. Then ${\Psi_1\ne0}$, ${\Psi_2=0}$, ${\Psi_3\ne0}$ determine \emph{type~I}.

In \cite{PodolskyPapajcik:2022} we identified the class of (cyclic symmetric) \emph{charged black hole electrostatic solutions}~\cite{Peldan:1993}, the 2+1 analogue to the Reissner--Nordstr\"om--(anti-)de~Sitter solution (Eq.~(192) of \cite{PodolskyPapajcik:2022} and Sec.~11.2 of \cite{Garcia:2017}). This arises as the special subcase ${Q=\hbox{const.}}$, ${e=0}$ of \eqref{RTmetric-final}. Such spacetimes are also of type~I, in agreement with Sec.~11.1.5 of \cite{Garcia:2017}. Interestingly, \emph{on the horizons}, localized by the condition ${H=0}$, the scalar ${\Psi_3}$ in \eqref{eq:null-rotatin-fixed-k-RT-EM} vanishes. So according to Tab.~\ref{Tab:Algorithm-special-case} these horizons are of \emph{type~III}. Eq.~\eqref{Class_eq} determining the CANDs for \eqref{eq:null-rotatin-fixed-k-RT-EM} becomes
\begin{equation}
( 1 + H\,K^2 )\, K=0\, .
\end{equation}
For ${H>0}$ (above the horizon) there are thus \emph{two complex CANDs}, so that this region is of algebraic subtype~I$_{\rm c}$. In contrast, for ${H<0}$  (below the horizon) there are four real CANDs, and thus the region is of subtype~I$_{\rm r}$.

\section*{Summary}

We have presented a new convenient method of algebraic classification for 2+1 geometries. It is independent on any field equations. The procedure has 5 simple steps:
\begin{enumerate}
\item Calculate the Cotton tensor $C_{abc}$ using \eqref{Cotton_Definition}.
\item Choose any null triad ${\{ \boldk, \, \boldl, \, \boldm \}}$ satisfying \eqref{null-comp}.
\item Evaluate the Cotton scalars~$\Psi_{\rm A}$ by \eqref{Psi}.
\item Calculate the invariants ${I, J, G, H, N}$ by \eqref{Invariants_IJGHN}.
\item Use the algorithm presented in Fig.~\ref{Flow_diagram} (or  Tab.~\ref{Tab:Algorithm-special-case}).
\end{enumerate}
The types I, II, III, N, D, O correspond to canonical forms of the Cotton scalars (Tab.~\ref{Tab-classification}), with specific multiplicity of the 4 CANDs (Tab.~\ref{Tab:algebraic-types}). Our method also suggests the refinement into the subtypes~I$_{\rm r}$, II$_{\rm r}$, D$_{\rm r}$ (all CANDs are real) and subtypes~I$_{\rm c}$, II$_{\rm c}$, D$_{\rm c}$ (complex CANDs). We proved that this agrees with previous classification scheme based on eigenvalues and Jordan forms of the Cotton--York tensor~${Y_a}^b$ (Tab.~\ref{Tab:Normal-Types}). As demonstrated in an explicit example (Sec.~\ref{sec:example}), it provides additional insight into the structure of exact spacetimes in 2+1 gravity.


This work has been supported by the Czech Science Foundation Grant No.~GA\v{C}R 23-05914S.


\begin{thebibliography}{10}


\bibitem{Will:2018}
C.~M.~Will,
{\it Theory and Experiment in Gravitational Physics}
(Cambridge University Press, Cambridge, England, 2018).

\bibitem{Rovelli:2010}
C.~Rovelli,
{\it Quantum Gravity}
(Cambridge University Press, Cambridge, England, 2010).

\bibitem{Carlip:2003}
S.~Carlip,
{\it Quantum Gravity in 2+1 Dimensions}
(Cambridge University Press, Cambridge, England, 2003).

\bibitem{Garcia:2017}
A.~A.~Garc\'ia-D\'iaz,
{\it Exact Solutions in Three-Dimensional Gravity}
(Cambridge University Press, Cambridge, England, 2017).

\bibitem{Stephanietal:2003}
H.~Stephani, D.~Kramer, M.~MacCallum, C.~Hoenselaers, and E.~Herlt,
{\it Exact Solutions of Einstein's Field Equations}
(Cambridge University Press, Cambridge, England, 2003).

\bibitem{NewmanPenrose:1962}
E.~Newman and R.~Penrose,
An approach to gravitational radiation by a method of spin coefficients,
{\it J.~Math.~Phys.} {\bf 3} (1962) 566; {\bf 4} (1963) 998.

\bibitem{PenroseRindler:1984}
R.~Penrose and W.~Rindler,
{\it Spinors and Space-Time}
(Cambridge University Press, Cambridge, England, 1984, 1986).

\bibitem{Coley:2008}
A.~Coley,
Classification of the Weyl tensor in higher dimensions and applications,
{\it Class. Quantum Grav.} {\bf 25} (2008) 033001.

\bibitem{OrtaggioPravdaPravdova:2013}
M.~Ortaggio, V.~Pravda and A.~Pravdov\'a,
Algebraic classification of higher dimensional spacetimes based on null alignment,
{\it Class. Quantum Grav.} {\bf 30} (2013) 013001.

\bibitem{KrtousPodolsky:2006}
P.~Krtou\v{s} and J.~Podolsk\'y,
Asymptotic structure of radiation in higher dimensions,
{\it Class. Quantum Grav.} {\bf 23} (2006) 1603.

\bibitem{Barrow}
J.~D.~Barrow, A.~B.~Burd, and D.~Lancaster,
Three-dimensional classical spacetimes,
{\it Class. Quant. Grav.} {\bf 3} (1986) 551.

\bibitem{GHHM}
A.~A.~Garc\'ia, F.~W.~Hehl, C.~Heinicke, and A.~Mac\'ias,
The Cotton tensor in Riemannian spacetimes,
{\it Class. Quant. Grav.} {\bf 21} (2004) 1099.

\bibitem{Cotton:1899}
\'E.~Cotton,
Sur les vari\'et\'es \`a trois dimensions,
{\it Annales de la facult\'e des sciences de Toulouse (II)} {\bf1} (1899) 385.

\bibitem{York}
J.~W.~York~Jr.,
Gravitational degrees of freedom and the initial-value problem,
Phys. Rev. Lett. {\bf 26} (1971) 1656.

\bibitem{PapajcikPodolsky:2023}
M.~Papaj\v{c}\'ik and J.~Podolsk\'y,
Algebraic classification of 2+1 geometries: a new approach,
{\it to be submitted}.

\bibitem{PodolskyPapajcik:2022}
J.~Podolsk\'y and M.~Papaj\v{c}\'ik,
All solutions of Einstein--Maxwell equations with a cosmological constant in 2+1 dimensions,
Phys. Rev. D {\bf 105} (2022) 064004.

\bibitem{Peldan:1993}
P.~Peldan, Unification of gravity and Yang-Mills theory in (2+1)-dimensions,
Nucl. Phys.~B {\bf 395} (1993) 239.


\end{thebibliography}
\end{document}